# Cooperative behavior of molecular motions giving rise to two glass transitions in the same supercooled mesophase of a smectogenic odd liquid crystal dimer

*David O. López[1], Josep Salud[1], María Rosario de la Fuente[2], Nerea Sebastián[3], Sergio Diez-Berart[1,*]*

[1]*Grup de Propietas Físiques dels Materials (GRPFM), Departament de Física, E.T.S.E.I.B. Universitat Politècnica de Catalunya. Diagonal 647, E- 08028 Barcelona, Spain*
[2]*Departamento de Física Aplicada II, Facultad de Ciencia y Tecnología, Universidad del País Vasco. Apartado 644, E-48080 Bilbao, Spain*
[3]*Jožef Stefan Institute, Jamova cesta 39, SI-1000 Ljubljana, Slovenia*
email: sergio.diez@upc.edu

In the present work, a detailed analysis of the glassy behavior and the relaxation dynamics of the liquid crystal dimer α-(4-cyanobiphenyl-4'-yloxy)-ω-(1-pyrenimine-benzylidene-4'-oxy) heptane (CBO7O.Py) throughout both nematic and smectic A mesophases by means of broadband dielectric spectroscopy has been performed. CBO7O.Py shows three different dielectric relaxation modes and two glass transition ($T_g$) temperatures: the higher $T_g$ is due to the freezing of the molecular motions responsible of the relaxation mode with the lowest frequency ($\mu_{1L}$); the lower $T_g$ is due to the motions responsible of the two relaxation modes with highest frequencies ($\mu_{1H}$ and $\mu_2$), which converge just at their corresponding $T_g$. It is shown how the three modes follow a critical-like description via the dynamic scaling model. The two modes with lowest frequencies ($\mu_{1L}$ and $\mu_{1H}$) are cooperative in the whole range of the mesophases, whereas the highest frequency mode ($\mu_2$) is cooperative just below some cross-over temperature. In terms of fragility, at the glass transition, the ensemble ($\mu_{1H}$and+$\mu_2$) presents a value of the steepness index and $\mu_{1L}$ a different one, meaning that fragility is a property intrinsic of the molecular motion itself. Finally, the steepness index seem to have a universal behavior with temperature for the dielectric relaxation modes of liquid crystal dimers, being almost constant at high temperatures and increasing drastically when cooling the compound down to the glass transition from a temperature about (¾)$T_{NI}$.

## I. INTRODUCTION

Understanding the molecular dynamics of a material is of primary importance from both theoretical and experimental points of view. Liquid crystal dimers, formed by two (semi)rigid units linked by a flexible spacer, are very good candidates for the study of molecular dynamics as they potentially present a rich and interesting variety of molecular motions in a same compound and these may be detected by dielectric spectroscopy [1-11]. We may refer to two kinds of motions: molecular, when the whole molecule reorients and intramolecular, when just a part of the molecule reorients.

As a consequence of steric interactions, molecular (and also intramolecular) motions can become cooperative. As a simple picture, we may say that the "free" volume for the molecules (or the parts of the molecule, if we deal with intramolecular motions) to rearrange gets smaller when cooling the material and, therefore, motions become cooperative among clusters of (parts of) molecules, the lower the temperature is, the greater the number of cooperative (parts of) molecules in these clusters is [12-14]. If the cooling rate of the material is fast enough, the moving part may not have time to rearrange and arrive to the equilibrium once the phase transition temperature to a more ordered phase is reached. If this is the case, it is said that the original disordered phase (ergodic state) is supercooled and it could happen, if temperature continues dropping, that vitrification to the glassy state (non-ergodic state) takes place [12-21] at the glass transition temperature, $T_g$. So, molecular cooperativeness and glass transition are closely related phenomena and the understanding of the former will help in deepening in the yet unexplained mysteries of the latter. Such an intramolecular coupling is strongly affected by the length of the linking chain. That is, the shorter the linking chain is, the larger the coupling is.

Some years ago, some of the authors of the present work studied the cooperativeness of these internal motions in a glass forming liquid crystal dimer 1'',7''-bis(4-cyanobiphenyl-4'-yl) heptane (CB7CB), in which the vitrified mesophase is the brand new twist-bend nematic phase, $N_{tb}$ [6-10]. This dimer, being symmetric, just presents two intermolecular dielectric relaxation modes [2,4,6,7], one at higher frequencies, $\mu_2$, which is due to precessions of the rigid cyanobiphenyl units around the mesophase director and the other one at lower frequencies, $\mu_1$, caused by the flip-flop reorientations of the rigid units around their short axes. Those molecular motions represented by both modes are strongly coupled when approaching the glass transition and it can be seen how they change in a coordinated fashion with temperature. Additionally, both motions seem to be responsible of the glassy dynamics with just one glass transition temperature and correspondingly one unique value of the fragility steepness index, *m*, at $T_g$ [22-24]. Such an index is defined as the absolute value of the slope of the log(τ) vs $T_g$/T, being τ the relaxation time related to the frequency of maximum dielectric loss of the mode. When *m* is low (16 is the lowest value, when the relaxation mode is Arrhenius) the material is defined as "strong" and the higher *m* is, the more "fragile" the material is. The $\mu_1$ mode seems to be cooperative in the entire temperature range of the mesophases (nematic, N, and $N_{tb}$), while $\mu_2$ is non-cooperative at high temperatures but becomes cooperative about 55 K above the glass transition. It is precisely from this temperature on that both modes behave in a very similar way, as if they were strongly coordinated. The so called dynamic scaling (DS) model seems to explain satisfactorily well the molecular dynamics of CB7CB. This model provides a critical-like description linked to dynamic domains with cooperative motions.

A very interesting result that we have recently presented is the presence of two close glass transition temperatures in the same supercooled mesophase. This result has been obtained by calorimetric measurements, dielectric spectroscopy and TSDC measurements in the series of non-symmetric liquid crystal dimers α-(4-cyanobiphenyl-4'-yloxy)-ω-(1-pyrenimine-benzylidene-4'-oxy) alkanes (CBOnO.Py) with n being an odd number [25,26]. The rigid units of these dimers are a cyanobiphenyl group and a pyrene group, and they were first synthesized with the idea of obtaining a liquid crystalline compound (as the cyanobiphenyl group is pro-mesogenic) which could vitrify (triggered by the bulky pyrene unit) [27]. In these dimers three dielectric relaxation modes have been identified [4,25,26,28,29]: the one at higher frequencies, $\mu_2$, due to precessions of the cyanobiphenyl rigid units around their long axes (the pyrene groups do also perform similar precessional motions, but they cannot be detected dielectrically); the ones at intermediate, $\mu_{1H}$, and lowest frequencies, $\mu_{1L}$, due to the flip-flop reorientations around their short axes of the cyanobiphenyl groups and the pyrene units, respectively. In these dimers, regardless of being nematogenic (n=11) or smectogenic (n=9), both $\mu_{1H}$ and $\mu_2$ converge at some temperature above the glass transition in one unique mode and are responsible of one $T_g$, whereas $\mu_{1L}$ vitrifies at a close but different temperature. This phenomenon is explained as the different thermal energy needed to activate the motions of the different rigid units of the dimer [25,26]. What is not so clear from the previous referred works is the nature of the intermolecular cooperativeness and intramolecular coupling of the molecular motions, as well as that of the fragility steepness index *m* at $T_g$. In the present paper we try to answer these questions with a detailed analysis of the dimer with n=7, CBO7O.Py. For this purpose, we follow an exhaustive methodology for the dielectric data treatment by examining different phenomenological descriptions for the glass transition and contrasting them with the experimental data. From here, we may partially understand the cooperativeness of the molecular motions in the entire temperature range from the N-I phase transition down to the glass transition. Regarding the fragility, which was first considered as a glass-forming material property [22-24] and later as a property of the glass-forming disordered phase for glassy states with just one $T_g$ [7], we shall expand this concept to a phase with several glass transition temperatures.

The work is divided as follows: we start describing the material and the dielectric spectroscopy measurements in the experimental section, we then present the results and the data analysis and the subsequent discussion and finally we come to the concluding remarks.

## II. EXPERIMENTAL DETAILS

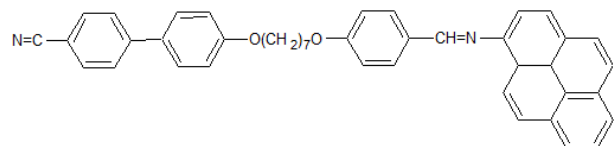

Scheme 1.- Molecular structure of CBO7O.Py.

The synthesis and purification of CBO7O.Py was performed according to the work of Attard et al. [27]. The phase sequence of the material on heating from room temperature is the following [29]:

Cr-428.2 K-N-433.4 K-I

If the sample is cooled down from the isotropic phase at 1 K·min$^{-1}$, the nematic phase is supercooled ($N_{sc}$) and the crystallization is avoided. The transition to a supercooled smectic A phase ($SmA_{sc}$) can be observed and, ultimately, this phase vitrifies ($SmA_g$). When heating up again from the glassy state, the phase sequence is [29]:

$SmA_g$-310.6 K-$SmA_{sc}$-333.5 K-$N_{sc}$-344 K-Cr-428.2 K-N-433.4 K-I

Dielectric spectroscopy measurements were performed with two equipments: HP 4291A impedance analyzer for frequencies from $10^6$ Hz to $10^9$ Hz and Alpha impedance analyzer from Novocontrol for frequencies from $10^{-3}$ Hz to $10^6$ Hz. The cell consists of two gold-plated brass electrodes of 5 mm of diameter separated by 50 μm thick silica spacers. The sample is held in a cryostat and the temperature control is performed by a System Quatro from Novocontrol. Additional details of the technique can be found elsewhere [20,30]. Dielectric measurements were performed on cooling and on heating with stabilization at different temperature steps and a temperature control of about 20 mK.

## III. RESULTS AND DATA ANALYSIS
### A. Relaxation times

The dielectric behaviour of CBO7O.Py at temperatures high above the glass transition has already been reported [29]. In the cited work, the identification of the different motions to the above cited three relaxation modes is made. On the other hand, in a more recent study, mainly by means of Thermal Stimulated Depolarization Currents (TSDC), the motions represented by the two dielectric relaxation modes at higher frequencies seemed to be frozen at one $T_g$, while the motions represented by the lowest frequency relaxation mode is frozen at a higher $T_g$ [26].

In the present paper the dielectric data are extended to low temperatures close to the glass transition, covering a broad range of temperatures from the isotropic phase.
Fig. 1 shows, as an example, the real and imaginary parts of the complex dielectric permittivity at 347 K and 319 K for planar (Fig. 1(a)) and homeotropic (Fig. 1(b)) alignments of the sample. Planar alignment is spontaneously achieved at the cells and a DC bias of 20 V must be applied to get the homeotropic one in the whole range of temperatures. Solid and dashed lines correspond to the fittings of experimental data to the empirical function:

$$\varepsilon(\omega) = \sum_k \frac{\Delta\varepsilon_k}{\left[1+(i\omega\tau_{k,HN})^{\alpha_k}\right]^{\beta_k}} + \varepsilon_\infty - i\frac{\sigma_{DC}}{\omega\varepsilon_0} \quad (1)$$

where k accounts for the relaxation modes present in the phase and each one is fitted according to the Havriliak-Negami function; $\Delta\varepsilon_k$ and $\tau_{k,HN}$ are the dielectric strength and the relaxation time (related to the frequency of maximum dielectric loss) of the modes, respectively; $\alpha_k$ and $\beta_k$ are parameters that describe the shape (width and symmetry) of the relaxation spectra; $\varepsilon_\infty$ is the dielectric permittivity at high frequencies (but lower than those corresponding to atomic and electronic resonance phenomena); and $\sigma_{DC}$ is the electric conductivity. In the quasi-planar alignment, the $\mu_2$ mode together with electric conductivity is clearly dominant, as can be observed in Fig. 1. The fitting parameters of $\mu_2$ are $\alpha<1$ and $\beta=1$ (Cole-Cole behavior), $\alpha$ ranging from 0.7 at high temperatures to 0.5 close to the glass transition. At 319 K, quite close to the glass transition, the $\mu_{1H}$ mode can be clearly observed. In the homeotropic alignment the $\mu_{1L}$ mode clearly dominates over the $\mu_{1H}$ mode, the $\mu_2$ mode having a very small strength. Both $\mu_{1L}$ and $\mu_{1H}$ are Debye-like ($\alpha=1$, $\beta=1$). At 347 K conductivity is also present. The subsequent data analysis is focused on, $\tau_{k,max}$, the inverse of the frequency of maximum dielectric loss of each relaxation mode, which is determined as

$$\tau_{k,max} = \tau_{k,HN}\left[\frac{\sin\frac{\pi\alpha_k\beta_k}{2\beta_k+2}}{\sin\frac{\pi\beta_k}{2\beta_k+2}}\right]^{1/\alpha_k}. \quad (2)$$

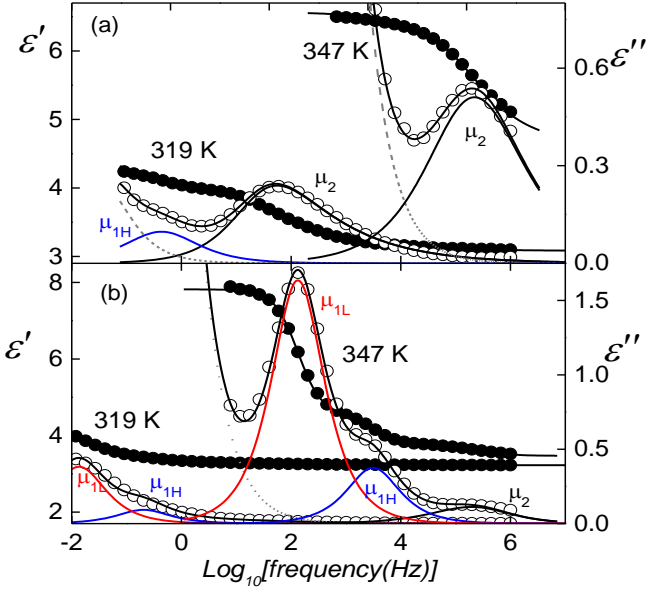

Fig. 1.- Frequency dependence of the complex dielectric permittivity of CBO7O.Py at 347 K (N) and 319 K (SmA) for (a) planar and (b) homeotropic alignment under 20 V of DC bias. Circles account for experimental real (full) and imaginary (empty) part. Fittings to eq. (1) are shown by the lines. Dotted line accounts for conductivity.

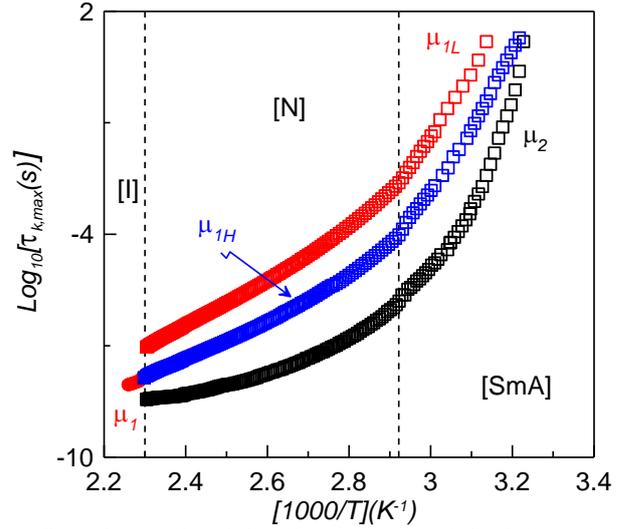

Fig. 2.- Arrhenius plot of the relaxation timess of the different dielectric modes.

The relaxation times $\tau_{k,max}$ are represented in an Arrhenius plot in Fig. 2. As it has been explained in a previous work [25,26], the mechanisms responsible for both $\mu_{1H}$ and $\mu_2$ are the (different) reorientations of the cyanobiphenyl group, which become indistinguishable at low temperatures close to $T_g$. In the other side, the $\mu_{1L}$ mode represents the reorientations of the pyrene group (which can be detected dielectrically because they imply a subsequent reorientation of the polar cyanobiphenyl group), bulkier than the cyanobiphenyl one and, therefore, easier to freeze at a higher $T_g$.

**B. Dynamic characterization**

The dielectric data analysis consists in trying to describe the cooperative behaviour of the relaxation modes, by fitting the temperature dependence of the relaxation time data ($\tau_{k,max}$) in the entire temperature range of the mesophases, down to the glass transition, to an adequate phenomenological model.

One of the most often used phenomenological expressions to describe the temperature dependence of the relaxation time is the Vogel-Fulcher-Tammann (VFT):

$$\tau = \tau_0 \exp\left[\frac{A}{T-T_0}\right], \quad (3)$$

where $\tau_0$ is the relaxation time in the high temperature limit, $A$ is an activation parameter and $T_0$ is the Vogel temperature. This equation establishes a divergence of the relaxation time at some finite temperature $T_0$ below the glass transition, but such a divergence has not been proven so far.

In order to test the adequacy of the VFT-model the temperature-derivative analysis [31-37] can be applied to eq. (3), which leads to

$$\left[\frac{d\ln\tau}{d(1/T)}\right]^{-1/2} = \left[\frac{H_A(T)}{R}\right]^{-1/2} = A^{-1/2}\left(1-\frac{T_0}{T}\right), \quad (4)$$

where $H_A(T)$ is the so-called *apparent activation enthalpy* [31]. The validity of the VFT-equation (eq. (3)) through eq.

(4) requires a linear dependence of our τ-data in a plot of $[H_A(T)/R]^{-1/2}$ as a function of the inverse temperature. Fig. 3 shows such a plot in which the three relaxation modes ($\mu_{1L}$, $\mu_{1H}$ and $\mu_2$) are represented. It can be observed how at high temperatures in the N phase the activation enthalpies of both $\mu_{1L}$ and $\mu_{1H}$ modes evolve horizontally, showing an Arrhenius-like behaviour, while that corresponding to the $\mu_2$ mode seems to fall abruptly from its value right after the $T_{NI}$. From a temperature around 380 K (about 2.6 scaled as 1000/T), the activation enthalpies corresponding to the three relaxation modes follow a linear (non-horizontal) trend down to the glass transition, verifying a VFT-like behaviour. In addition, it seems that the activation enthalpies corresponding to $\mu_{1L}$ and $\mu_{1H}$, merge at about 380 K and, at lower temperatures close to the N-SmA phase transition, they also merge with that of $\mu_2$. It is also noted that for the three modes the dynamics become more Arrhenius-like when temperature approaches the glass transition, as observed in other materials [7,38]. The fitting procedure is as follows: in a first step, $A$ and $T_0$ are obtained from the fitting of the experimental data to eq. (4) and, afterwards, the value of $\tau_0$ is obtained from the fitting of the data to eq. (3) to obtain the pre-factor $\tau_0$. Two different strategies have been followed for fitting the data to eq. (4): in a first approach, one fitting is done for both $\mu_{1L}$ and $\mu_{1H}$ and another one for $\mu_2$ (dashed lines in Fig. 3), both from 380 K (about 2.6 scaled as 1000/T) to the lowest temperature. The fitting parameters are listed in Table I. Such a procedure leads to three different glass transition temperatures, which disagrees with those results from TSDC experiments, where just two glass transitions have been reported [26]. A second approach consists in fitting both $\mu_{1H}$ and $\mu_{1L}$ modes only in the SmA phase, while the $\mu_2$ mode is fitted as before. These alternative fittings of the curves corresponding to the two lowest modes in frequency are represented by continuous lines in Fig. 3 and its insets. The new $A$ and $T_0$ parameters after refining through eq. (3) to obtain the new pre-factor $\tau_0$ are also listed in Table I (with a [SmA] label). Under this second approach, even if three glass transition temperatures are also present, the difference between the two lowest is smaller than 1 K. Limiting the fittings of the two low frequency modes to the SmA phase, the obtained parameters are closer to the experimental values of the glass transition temperatures. The inset in Fig. 3 shows the relaxation data for the three modes in an Arrhenius plot with the VFT-fittings, according to the two investigated approaches. It must be said that separate fittings for the $\mu_{1L}$ and $\mu_{1H}$ modes have also been tested, but there is no improvement in the results.

Anyway, the observed ambiguity in the data analysis, depending on the fitting range, indicates that we should be the careful about the adequacy of the VFT-equation, the validity of which has in fact been called into question once and again [38-41].

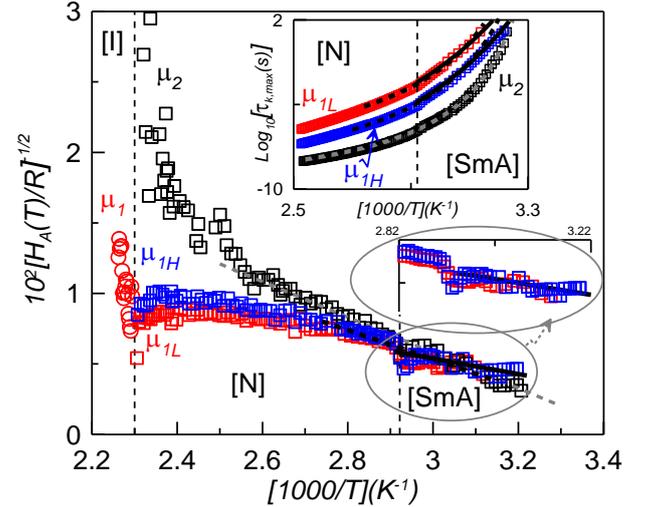

Fig. 3.- Results of the temperature-derivative analysis [eq. (4)] applied to the three relaxation modes of CBO7O.Py. Linear dependences indicate domains of validity of the VFT-model. The inset presents the relaxation time data as an Arrhenius plot. With respect to both $\mu_{1L}$ and $\mu_{1H}$ modes, dashed lines account for the fitting including N and SmA phases, whereas solid lines account for the fitting just at the SmA phase.

In addition to the broadly used VFT-model, there are several phenomenological models based in the link between the glass phenomenon and the increasing cooperativeness of molecular motions when approaching $T_g$ which can be also considered. One of these alternative models, introducing a mean-field description, is the so-called dynamic scaling (DS) one [14,42,43], for which

$$\tau = \tau_0 \left[ \frac{T - T_C^{DS}}{T_C^{DS}} \right]^{-\phi} \qquad (5)$$

where $T_C^{DS}$ is the temperature of a virtual phase transition below $T_g$ (somehow comparable to the Vogel temperature $T_0$ in VFT), also known as the critical temperature; $\tau_0$ is the relaxation time at $2T_C^{DS}$ and $\phi$ is the critical exponent, which has a value about 9 for glass forming polymers [14] but that may vary between 6 and 15 for other glass forming systems [31], including liquid crystals [7,31,44]. The DS model is valid at temperatures in the vicinity of $T_g$, below a certain caging temperature, $T_A$, above which there is no cooperative motions of the entities. For the high temperature domain, above $T_A$, where the coupling mechanisms can be disregarded, it seems to be adequate the description provided by the Mode Coupling Theory (MCT). Equation for MCT is similar to that of DS model (eq. (5)):

| Mode | $Log_{10}[\tau_0(s)]$ | $A(K)$ | $T_0(K)$ | $T_g(K)$ | Range[$1000/T(K^{-1})$] | $\chi^2$ |
|---|---|---|---|---|---|---|
| $\mu_{1,L}$ | -8.23 | 760.90 | 284.27 | 316.57 | 2.74-3.15 | 0.009 |
|  | -12.43 | 2094.6 | 251.28 | 314.32 | 2.92-3.2 [SmA] | 0.009 |
| $\mu_{1,H}$ | -9.63 | 760.90 | 284.27 | 312.68 | 2.74-3.15 | 0.05 |
|  | -13.83 | 2094.6 | 251.28 | 308.71 | 2.92-3.2[SmA] | 0.003 |
| $\mu_2$ | -10.08 | 526.64 | 288.99 | 307.92 | 2.54-3.2 | 0.003 |

TABLE I.- Fitting parameters according to eqs. (3) and (4) for the different dynamic domains and the calculated glass transition temperature for the $\mu_{1,L}$, $\mu_{1,H}$ - and- $\mu_2$-modes.

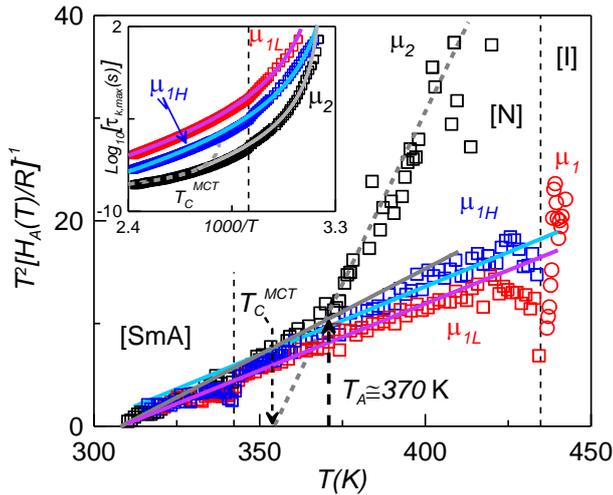

Fig. 4.- Results of the temperature-derivative analysis [eq. (7)] applied to the three relaxation modes of CBO7O.Py. Linear dependences indicate domains of validity of the critical-like description. The inset presents the relaxation time data as an Arrhenius plot. Solid and dashed lines correspond to the DS and MCT descriptions, respectively.

$$\tau = \tau_0 \left[\frac{T - T_C^{MCT}}{T_C^{MCT}}\right]^{-\phi} \quad (6)$$

but $T_C^{MCT}$ now accounts for the crossover temperature from the ergodic to the non-ergodic domain and seems to be correlated with $T_A$, $\tau_0$ is the relaxation time at the limit of high temperatures and the critical exponent ranges from 1.5 to 4 [44]. When applying the temperature-derivative analysis to the critical-like description of both DS and MCT models, we get the following expression [35]:

$$T^2 \left[\frac{d \ln \tau}{d(1/T)}\right]^{-1} = \left[\frac{T^2 R}{H_A(T)}\right] = \frac{1}{\phi}(T - T_C^X), \quad (7)$$

where X accounts for both DS and MCT models. In such a case, the adequacy of our τ-data to both models requires a linear trend when $[T^2R/H_A(T)]$ is plotted against temperature. Fig. 4 shows such a plot for the three relaxation modes ($\mu_{1L}$, $\mu_{1H}$ and $\mu_2$). At first glance, both $\mu_{1L}$ and $\mu_{1H}$ modes exhibit a linear behaviour according to the DS-model over the SmA and N phases, up to the N-I phase transition. However, $\mu_2$-mode clearly shows two linear trends, one at low temperatures from $T_g$ up to about 370 K (DS-model) and the other at high temperatures according to the MCT description. The linear fittings according to eq. (7) provide preliminary values of $\phi$, $T_C^{DS}$ and $T_C^{MCT}$ for each mode and dynamic description, either DS or MCT. The final fitting parameters are obtained to either eq. (5) or eq. (6) with the pre-factor $\tau_0$ dashed line (MCT-description). The inset shows the and are listed in Table II. In Fig. 4 the fittings to eq. (7) are represented by continuous lines (DS-description) and a relaxation time data as an Arrhenius plot also with the fitting-curves according to eqs. (5) and (6).

One of the most important results is the adequacy of the critical-like description (DS and MCT models) to describe the relaxation time data for CBO7O.Py. First of all, two glass transitions are obtained, one related to $\mu_{1H}$ and $\mu_2$ (according to Table II, the difference in the independent fittings is very small, ~0.2 K, and so is considered to be the same), and the other related to $\mu_{1L}$, about 6 K higher, as predicted from TSDC experiments [26]. The inverse of the exponent $\phi$ is the slope of the represented lines in Fig. 4. The DS and MCT exponents are ranged in the typical values for liquid crystals [7] and the caging temperature $T_A$ is about $1.04 T_C^{MCT}$ as expected [5].

The caging temperature $T_A$ deserves a special mention. According to the critical-like description, $T_A$ was postulated as the temperature above which intermolecular cooperativeness is lost, i.e., molecules are able to move freely [45]. This simplified picture applied to the liquid crystal dimer CBO7O.Py, in which several molecular motions are identified, requires to reformulate the concept of cooperativeness as an inter or intramolecular coupling of such motions. It is clearly observed how those motions identified with the $\mu_{1H}$ and $\mu_{1L}$ modes follow the DS-model from glass transition up to the N-I phase transition and then, they should be thought as cooperative in the entire range of both SmA and N phases. It should be remembered that such motions correspond to the flip-flop of the terminal rigid units, highly coupled in the anisotropic environment of the SmA and N phases. It is clear that these flip-flop motions require a high available volume to take place and, therefore, they should present intermolecular cooperativeness. The intramolecular coupling between the rigid units of the molecule also affects the motions, as their strengths are coordinated [4,28,29]. It is quite feasible that this intramolecular coupling also translates in a cooperativeness that should be added to the intermolecular one, but probably to a much lesser extent.

Nevertheless, those motions represented by the $\mu_2$ dielectric relaxation mode, attributed to precessional motions of the dipolar groups of the rigid cyanobiphenyl units about the mesophase director, require a much lower available volume to complete. It does not seem likely, therefore, that these motions are cooperative in an intermolecular way but at very low temperatures, close to the glass transition. Indeed, they are very influenced at such low temperatures by the highly-coupled flip-flop motions and the anisotropic environment (SmA phase and extended conformers), being highly cooperative. It may happen that, as well as for the flip-flop motions, both inter- and intramolecular cooperativeness are present, the weight of the intramolecular cooperativeness being higher for the precessional motions. These lose all kind of cooperativeness at $T_A$ (370 K or 1.04 $T_C^{MCT}$ ). This behaviour was already observed for the symmetric dimer CB7CB (a so called twist-bent dimer) where only two different motions were identified [7].

| Mode | $Log_{10}[\tau_0(s)]$ | $T_C^X$ (K) | $\phi$ | $T_g(K)$ | Range[$1000/T(K^{-1})$] | Description | $\chi^2$ |
|---|---|---|---|---|---|---|---|
| $\mu_{1,L}$ | -9.87 | 308.15 | 7.7 | 317.0 | 2.4-3.2(a) | DS | 0.004 |
| $\mu_{1,H}$ | -10.71 | 304.12 | 7.6 | 310.7 | 2.4-3.2(a) | DS | 0.02 |
| $\mu_2$ | -9.39 | 356.7 | 1.5 | --- | 2.4-2.7 | MCT | 0.0005 |
|  | -11.63 | 308.59 | 6.2 | 310.5 | 2.7-3.2 | DS | 0.009 |

TABLE II.- Fitting parameters according to eqs. (6) and (7) for the different dynamic domains and the calculated glass transition temperature for the $\mu_{1,L}$, $\mu_{1,H}$- and- $\mu_2$, -modes.

### C. Fragility

The ability of a material to form a glass is related with the so-called fragility concept [22-24] that accounts for the way in which the dynamic properties change as the material approaches the glass transition. In complex materials like liquid crystal dimers, instead to use the *m*-fragility, it results more convenient the temperature dependent steepness index [35], defined as

$$m(T) = \frac{H_A(T)\log e}{RT}. \quad (8)$$

When taking T=$T_g$, we have the *m*-fragility or $m(T_g)$ [23,24,46-48].

The differential mark of the LC dimer with respect to any other molecular liquid or mesogen lies in the large complexity of the former, which reflects in the behaviour of the steepness index with temperature. Fig. 5 shows *m(T)* as a function of temperature for the three relaxation modes $\mu_{1L}$, $\mu_{1H}$ and $\mu_2$ of CBO7O.Py. By combining eqs. (7) and (8), we obtain

$$m(T) = \phi \log e \frac{T}{T - T_C^X} \quad (9)$$

According to eq.(8), if $m(T)^{-1}$ is plotted against inverse temperature a linear dependence, as shown in the inset of Fig. 5, is obtained. Eq. (9) also points out an asymptotic behaviour of the m(T) at $T_C^{DS}$, i.e. at temperatures slightly below the glass transition temperature (see table 2; about 305 K for both $\mu_{1H}$ and $\mu_2$ and about 308 K for $\mu_{1L}$), fact that is clearly observed in Fig. 5 when m(T) data approaches the glass transition. On the other hand, at high temperatures m(T) tends to a constant value, different for each relaxation mode being $m_{\mu2}<m_{\mu1H}<m_{\mu1L}$. According to eq. (9), *m(T)* at high temperatures should be related with the critical exponent $\phi$ that yields $\phi_{\mu2}<\phi_{1H}<\phi_{1L}$ (see table 2). According to the observed variation of m(T) as temperature decreases (see Fig. 5) is quite complicated to give a precise value for the m-fragilty ($m(T_g)$). Two values of $m(T_g)$ can be inferred from Fig. 5, one for the $\mu_{1L}$ mode of 94 and another for both the $\mu_{1H}$ and $\mu_2$ modes of about 180. Due to the asymptotic behaviour at temperatures very close to the glass transition, there is a large difference between both m-values.

In a previous study [29], $m(T_g)$ for CBO7O.Py was obtained from heat capacity measurements through an interesting relationship introduced by Huang and McKenna for glass-forming polymers [49]:

$$m = 254 - 120 \frac{C_{p,sc}}{C_{p,g}}. \quad (10)$$

where $C_{p,sc}$ and $C_{p,g}$ are, respectively, the heat capacities of the supercooled liquid state and the corresponding glassy state at $T_g$. In ref. [29] the supercooled phase (sc) is identified with the SmA phase and taking into account the ratio ($C_{p,SmA}/C_{p,g}$) from heat capacity measurements we obtained $m(T_g)$ of about 100. The observed jump of heat capacity at the glass transition is mainly due to the bulkier pyrene group and, correspondingly, to the $\mu_{1L}$ relaxation mode [25,26]. Such a value of 100 is in reasonably agreement to the one obtained by dielectric measurements and the DS model for this particular mode ($m(T_g)$=94).

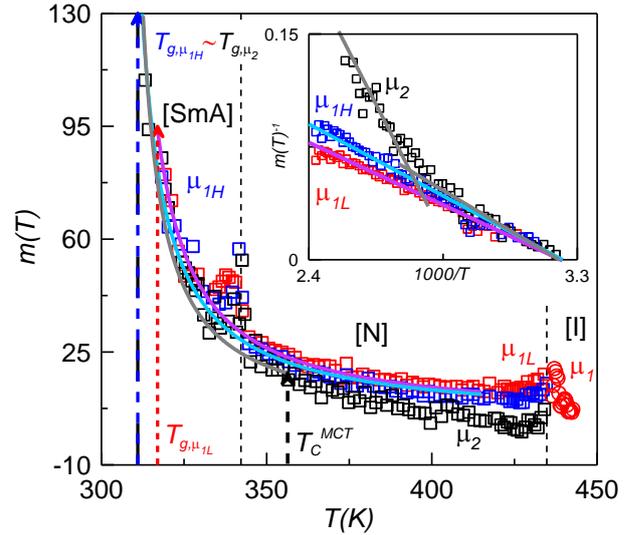

Fig. 5.- Results for the temperature dependent steepness index m(T) for the three relaxation modes of CBO7O.Py. Lines correspond to the m(T) functions according to the DS-model using the parameters given in table 2. The inset shows the linear dependence of the inverse of m(T) with the inverse of temperature. For the $\mu_2$ mode, both temperature regimes, consistent with DS and MCT-models are represented by both lines.

The information of Fig. 5 about the temperature dependent steepness index m(T) is really intriguing. At a certain temperature about 350 K, the m(T) associated to the different motions experience a sudden and continuous increase as temperature decreases. Could it be considered a common behaviour in liquid crystalline materials? Unfortunately, such an exhaustive analysis is uncommon and, at this moment, we only have data for another liquid crystal dimer, the symmetric twist-bend dimer CB7CB, with clear differences with the non-symmetric CBO7O.Py but with some common particularities. First of all, if we compare the mesophase sequence of CB7CB and CBO7O.Py, both compounds are very different: CB7CB exhibits two nematic mesophases, one of them being the new discovered twist-bend nematic phase which vitrifies by slow cooling at 276 K [7]. In the case of CBO7O.Py, only one nematic phase followed by a smectic A phase is observed when temperature decreases. By slow cooling SmA also vitrifies, but at two temperatures spanned 6 K (311 K and 317 K). The N-I phase transition temperatures are very different among them, about 390 K for CB7CB [7] and 433.4 K for CBO7O.Py [29]. However, regarding the molecular motions, they have more similarities: in CB7CB two different motions are distinguished, the flip flop of the rigid units of the molecules (called as $\mu_1$) and the precessional motions of the dipolar groups of the rigid units about the local director (called as $\mu_2$); in CBO7O.Py three molecular motions are identified, two of them linked with the flip flop of each of the rigid units ($\mu_{1,L}$, $\mu_{1,H}$) and the other one due to the precessional motions

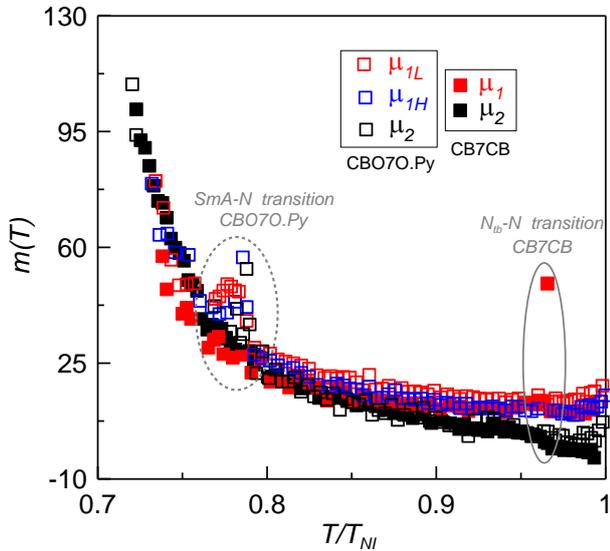

Fig. 6.- Results for the steepness index $m(T)$ for the relaxation modes of CBO7O.Py and CB7CB.

of the cyanobiphenyl group, $\mu_2$.

Fig. 6 shows data of $m(T)$ for both CBO7O.Py and CB7CB as a function of $(T/T_{NI})$. It is really very interesting to observe how the $m(T)$ related to the relaxation modes identified for both compounds, even being slightly different a high temperatures, merge at about 0.85 in $(T/T_{NI})$ and at slightly lower temperatures (at about 0.75 in $(T/T_{NI})$) experience an abrupt and coordinated increase as temperature decreases approaching the glass transition, irrespective the compound or molecular motion.

## IV. CONCLUDING REMARKS

In the present work we have analysed the glassy behaviour and the relaxation dynamics of the highly non-symmetric odd smectogenic dimer CBO7O.Py, formed by two (semi)rigid units, pyrene and cyanobiphenyl, linked via a flexible alkyloxy chain. This compound shows three relaxation modes in both N and SmA mesophases. The one at lower frequencies ($\mu_{1L}$) is due to end-over-end rotations of the bulky pyrene group, the one at intermediate frequencies ($\mu_{1H}$) to flip-flop motions of the cyanobihenyl group and the one at higher frequencies ($\mu_2$) due to precessions of the cyanobiphenyl group around the nematic director as well as rotations of the same unit around its large axis. These three modes can be frozen and they give rise to two different glass transition temperatures: $\mu_{1H}$ and $\mu_2$ merge at the glass transition, as they are due to the same rigid unit, whereas $\mu_{1L}$ freezes at a higher temperature, as the pyrene group is bulkier than the cyanobiphenyl [25,26]. The simultaneous presence of different relaxation modes with different glass transition temperatures in a same glass-forming phase, makes us reconsider the nature of the glass transition, which was first considered as a glass-forming material property and later as a property of the glass-forming disordered phase for glassy states.

The dynamic characterization of the data analysis has been made via two different models: Vogel-Fulcher-Tammann (VFT) and the critical-like description that combines the Dynamic Scaling (DS) model and the Mode Coupling Theory (MCT). This critical-like description reproduces better the experimental data and from such a model we have shown how both $\mu_{1L}$ and $\mu_{1H}$ modes are cooperative in the entire temperature range of the mesophases (they are described by the DS model) while the higher frequency $\mu_2$ mode is non-cooperative (well described by the MCT) above some cross-over temperature, the so-called caging temperature $T_A$, and below that temperature it is cooperative (following the DS model).

In terms of fragility we have proven how this property is intrinsic of the molecular motion, dielectrically represented by the relaxation mode and neither of the material itself, nor of the glass forming phase. When arriving to the glass transition, the steepness indices $m(T)$ of both $\mu_{1H}$ and $\mu_2$ become similar as the modes merge into one unique complex mode. As the $\mu_{1L}$ vitrifies at a higher temperature as that for the ensemble $\mu_{1H}+\mu_2$ mode, and taking into account that: 1) the steepness index vs T is asymptotic when arriving to the glass transition and 2) the ensemble $\mu_{1H}+\mu_2$ mode is presumably more complex than $\mu_{1L}$ and its $m$-fragility higher than that of $\mu_{1L}$. The value of the m-fragility for the $\mu_{1L}$ mode agrees reasonably well the relationship proposed by Huang and McKenna between this dynamic fragility and the thermodynamic fragility based in the ratio between the heat capacity of the supercooled and the glassy states at $T_g$.

Finally, it must be stated that the dependence of the steepness index $m(T)$ with temperature seems to be common for other liquid crystal dimers. At high temperatures, from the $T_{NI}$, the value of $m(T)$ does not change significantly. When arriving at a about $(3/4)T_{NI}$ the steepness index increases drastically when cooling the compound down to the glass transition.

## ACKNOWLEDGEMENTS

The authors are grateful for financial support from the MICINN project MAT2015-66208-C3-2-P (MINECO-FEDER). The authors also acknowledge the recognition from the Generalitat de Catalunya of GRPFM as consolidate research group (2014 SGR-190). N. S. thanks the "EU Horizon 2020 Framework Programme for Research and Innovation" for its support through the Marie Curie Individual Fellowship No. 701558.